\documentclass[pdflatex,sn-aps]{sn-jnl}

\usepackage{graphicx}%
\usepackage{xcolor}
\usepackage{multirow}%
\usepackage{amsmath,amssymb,amsfonts}%
\usepackage{amsthm}%
\usepackage{mathrsfs}%
\usepackage[title]{appendix}%
\usepackage{textcomp}%
\usepackage{manyfoot}%
\usepackage{booktabs}%
\usepackage{algorithm}%
\usepackage{algorithmicx}%
\usepackage{algpseudocode}%
\usepackage{listings}%
\usepackage{mhchem}
\usepackage{ragged2e}
\usepackage[normalem]{ulem} 
\usepackage{multirow}

\usepackage{siunitx}

\usepackage{caption}
\usepackage{subcaption}
\usepackage{array}
\usepackage{multirow}

\theoremstyle{thmstyleone}%
\theoremstyle{thmstyletwo}%

\theoremstyle{thmstylethree}%

\newgeometry{top=2.5cm, bottom=2.5cm, left=2.5cm, right=2.5cm}

\begin{document}

\title[First-Principles Investigation of the {ScSe2} TMDs Monolayer:
Structural Stability, Electronic and Optical Properties]{First-Principles Investigation of the {ScSe2} TMDs Monolayer:
Structural Stability, Electronic and Optical Properties}


\author*[1]{\fnm{Takwa} \sur{Ben Othman}}\email{benothmanetakwa960@gmail.com}
\equalcont{These authors contributed equally to this work.}

\author*[1]{\fnm{Mohsen} \sur{Jema\"i}}\email{mohsen.jemai@fst.utm.tn}
\equalcont{These authors contributed equally to this work.}

\author*[1]{\fnm{Said} \sur{Ridene}}\email{said.ridene@fst.utm.tn}
\equalcont{These authors contributed equally to this work.}

\affil*[1]{\orgdiv{Advanced Materials and Quantum Phenomena Laboratory,}, \orgname{Department of Physics, Sciences Faculty of Tunis, Tunis El Manar University}, \orgaddress{\street{Campus Universitaire El-Manar}, \city{Tunis}, \postcode{2092}, \state{Tunis}, \country{Tunisia}}}

%


\abstract{We present an ab initio investigation of the {ScSe2} monolayer using density functional theory (DFT), as employed in the CP2K code, to assess its structural stability and electronic properties. 
The electronic behavior of both phases is governed by the local coordination environment of scandium and the associated lifting of degeneracy of its $3d$ orbitals. The valence bands (VBs) are dominated by {Se}$-4p$ states in both phases, while the states near the Fermi level distinguish the metallic 1T phase from the semiconducting 2H phase. 
Optical response calculations reveal markedly different spectral signatures for the two polymorphs. Two findings emerge from this study: a polymorphism-driven metal-to-semiconductor transition without any change in chemical composition and distinct optical responses relevant to phase-selective optoelectronic devices.}



\maketitle

\section{Introduction}

The transition metal dichalcogenides (TMDs), of general formula MX$_2$ (M: transition metal; X: chalcogen, typically S, Se, or Te), form a particularly rich class encompassing both semiconducting and metallic compounds \cite{Rasmussen2015, Rabiaa2026}. These materials consist of X--M--X layers detained together by van der Waals forces \cite{Mohsen2023, Mohsen2024}, which allows mechanical or chemical exfoliation down to the monolayer limit. They crystallize mainly in two hexagonal polytypes: the 2H phase, with trigonal-prismatic coordination of the metal atom, typically semiconducting for group III, IV, and VI metals; and the 1T phase, with octahedral coordination, which can be metallic, semiconducting, or topologically nontrivial depending on the metal's $d$-orbital filling.

Scandium diselenide (ScSe2) occupies an unusual and still largely unexplored position within the TMD family. Scandium (Sc, $Z = 21$) is a group III transition metal with electronic configuration [Ar]$3d^{1}4s^{2}$. With a single $d$ electron, {ScSe2} exhibits a fundamentally different band-filling behavior from the extensively studied group VI TMDs (Mo, W) \cite{Rabiaa2026}. While DFT has successfully characterized analogous systems such as {SnSe2}, {WSe2}, and Sc-based MXene compounds for applications in batteries, optoelectronics, and photocatalysis \cite{EsSmairi2025}, systematic ab initio investigations of pristine {ScSe2} monolayers remain absent from the literature. The present study addresses this gap by employing DFT to evaluate the structural robustness, dynamical stability, and electronic band structure of {ScSe2} monolayers, with the aim of assessing their suitability for next-generation nanotechnological applications.

This work presents an exploratory ab initio DFT study of monolayer {ScSe2}, aiming to predict its structural and electronic properties by incorporating standard approaches for 2D materials: geometric optimization of equilibrium lattice parameters and bond lengths under symmetry constraints, calculation of vibrational frequencies to confirm the absence of imaginary modes and establish dynamical stability, and an in-depth analysis of the electronic band structure to determine the nature of the gap (direct or indirect) and its width, the dispersion of VB and conduction bands (CB), the role of atomic orbitals in shaping the electronic states, and the density of states (DOS) used to identify the respective contributions of scandium and selenium orbitals. This approach enables a quantitative assessment of the suitability of {ScSe2} for nanotechnology applications, building on theoretical frameworks validated for analogous systems such as {SnSe2} nanosheets, for which DFT has already established structure--activity relationships relevant to catalytic activity \cite{EsSmairi2025}. The overall goal is to highlight the potential of monolayer {ScSe2} for future applications in nanoelectronics and optoelectronics; a solid understanding of its fundamental properties is essential to guide both experimental work and functional-material development.

The manuscript is ordered as follows. Section~\ref{sec:methods} gives the theoretical foundations of DFT and the computational parameters used in this study. Section~\ref{sec:results} examines the structural characteristics and electronic properties of monolayer {ScSe2}, including a comparison of the LDA and GGA-PBE functional, the DOS on the band gap.  
Section~\ref{sec:optical} discusses the dielectric and optical properties of the two phases. The manuscript concludes with a summary of the main results and a discussion of potential applications and future research directions.

\section{Computational Methods}
\label{sec:methods}

The {ScSe2} monolayer is modeled within a three-dimensional supercell subject to periodic boundary conditions along all three axes. A vacuum gap of \SI{20}{\angstrom} is inserted along the out-of-plane $z$-direction to suppress interactions between periodic images; this value was validated by convergence tests on comparable TMD systems. 
The computational parameters used throughout this study are assembled in Table~\ref{tab:params}.

\begin{table}[htbp]
\centering
\caption{Computational parameters used for the DFT study of the {ScSe2} monolayer.}
\label{tab:params}
  \begin{tabular*}{1.0\textwidth}{@{\extracolsep{\fill}}ll}
\hline
Parameter & Adopted value \\
\hline
Exchange--correlation functional & PBE (structural) 
 /PBE or LDA (electronic) \\
Basis set (Sc, Se) & DZVP-MOLOPT-SR-GTH \\
Pseudopotential & GTH-PBE or LDA (11 $e^-$  
  for Sc; and 6 $e^-$ for Se) \\
Plane-wave cutoff & 600--800 Ry \\
Relative cutoff & 60 Ry \\
Vacuum spacing ($z$) & \SI{20}{\angstrom} \\
SCF convergence & $10^{-6}$ Ha ($1\,\text{Ha} = \SI{27.21}{eV}$) \\
Force convergence & $4.5\times 10^{-4}$ Ha/bohr \\
$k$-point grid (structural relaxation) & $8\times 8\times 1$ (Monkhorst--Pack) \\
$k$-point grid (electronic structure) & $12\times 12\times 1$ \\
\hline
\end{tabular*}
\end{table}

The geometry structure optimization has been achieved with the Broyden--Fletcher--Goldfarb--Shanno (BFGS) algorithm as employed in CP2K, converging simultaneously on the atomic forces and on the stress tensor to optimize the in-plane lattice parameters \cite{Fletcher1987}. Electronic band structures are computed along the standard high-symmetry path of the hexagonal Brillouin zone, $\Gamma \rightarrow M \rightarrow K \rightarrow \Gamma$ (or equivalently $M \rightarrow \Gamma \rightarrow K \rightarrow M$) \cite{Setyawan2010}, which also provides an estimate of the effective masses at the zone center ($\Gamma$).

\section{Structural and Electronic Properties}
\label{sec:results}

Two-dimensional TMDs have become a reference system for the study of physical phenomena specific to low-dimensional materials. While group VI TMDs such as {MoS2}, {WSe2}, {WS2} have been extensively characterized, group IIIB compounds containing scandium remain comparatively understudied, despite their promising potential for property tuning via phase polymorphism \cite{Cappelluti2013,Wang2012}.

We report the results of an ab initio DFT study of the {ScSe2} monolayer in its two polymorphic structures: the 1T phase, with octahedral coordination (space group $P\bar{3}m1$), and the 2H phase, with trigonal-prismatic coordination. All calculations were performed with two exchange--correlation functionals, the local density approximation (LDA) and the generalized-gradient approximation (GGA-PBE), including van der Waals corrections, in order to assess the sensitivity of the computed properties to the choice of exchange--correlation treatment. The analysis proceeds in four complementary steps. First, the electronic band structure is examined along the principal symmetry directions of the hexagonal Brillouin zone ($M \rightarrow \Gamma \rightarrow K \rightarrow M$), revealing a clear contrast between the metallic 1T phase and the indirect-gap semiconducting 2H phase. Second, the total and projected densities of states (TDOS/PDOS) are analyzed and related to the differentiated lifting of the Sc $3d$ orbital degeneracy under octahedral and prismatic crystal fields. 
Finally, the dielectric and optical properties of the {ScSe2} monolayer are computed from the complex dielectric function $\varepsilon(\omega) = \varepsilon_1(\omega) + i\varepsilon_2(\omega)$, obtained within linear-response theory 
establishing a connection between the electronic structure and spectroscopically observable signatures \cite{Dronskowski1993,Bader1990}.

\subsection{Band structure of the 1T phase}

The 1T phase adopts a {CdI2}-type structure with octahedral ($O_h$) coordination around the scandium atom. The space group is $P\bar{3}m1$ (No.\ 164), and the primitive hexagonal unit cell contains one formula unit, with atomic positions $\mathrm{Sc}(0,0,0)$, $\mathrm{Se}(1/3,2/3,z)$, and $\mathrm{Se}(2/3,1/3,-z)$, with $z \approx 0.25$ in fractional coordinates \cite{Wang2012,Ataca2012}.

\begin{figure}[h]
\centering
\begin{subfigure}[b]{0.75\textwidth}
\includegraphics[width=\textwidth]{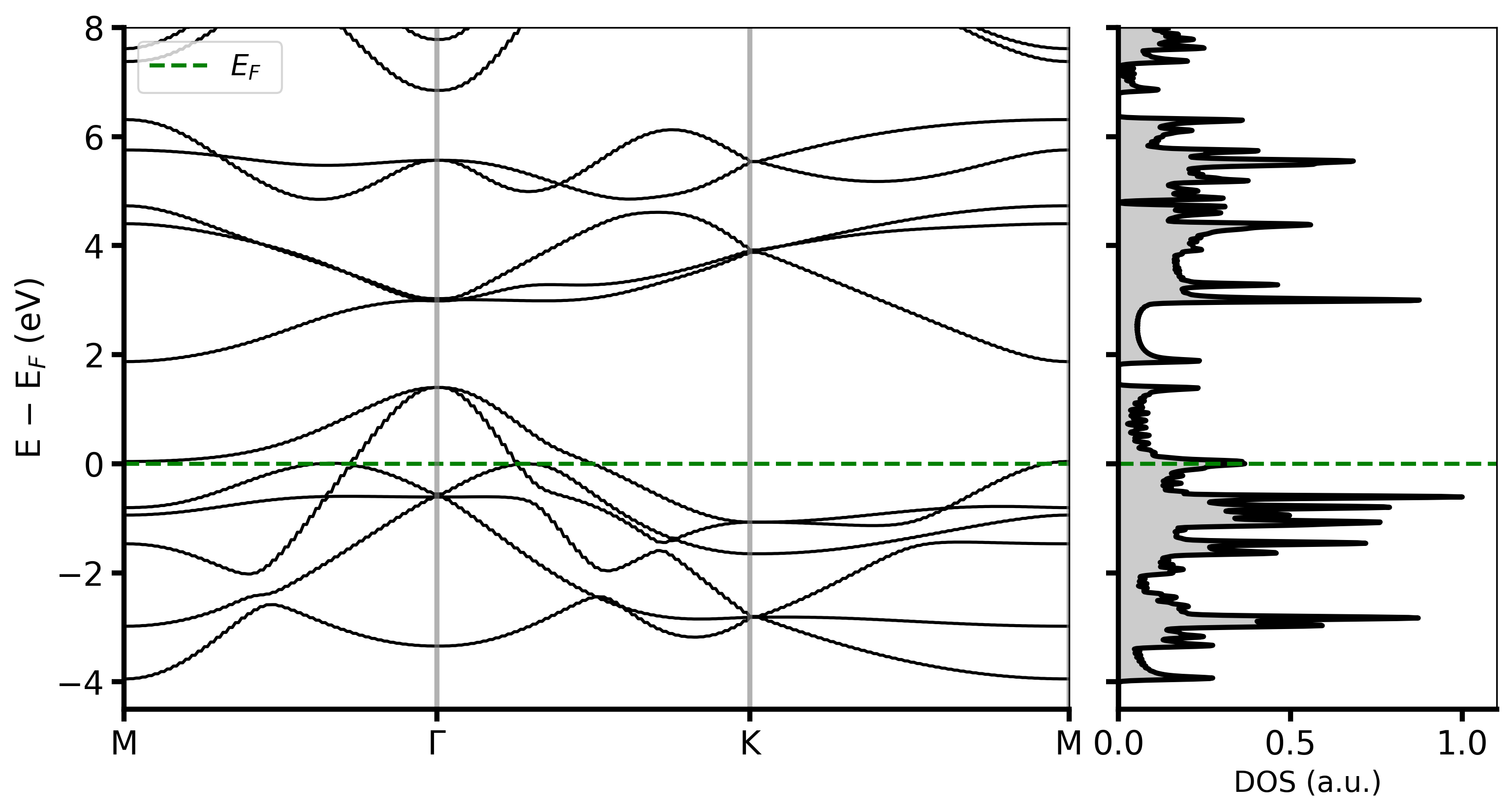}
\caption{LDA approximation}
\end{subfigure}\\[6pt]
\begin{subfigure}[b]{0.75\textwidth}
\includegraphics[width=\textwidth]{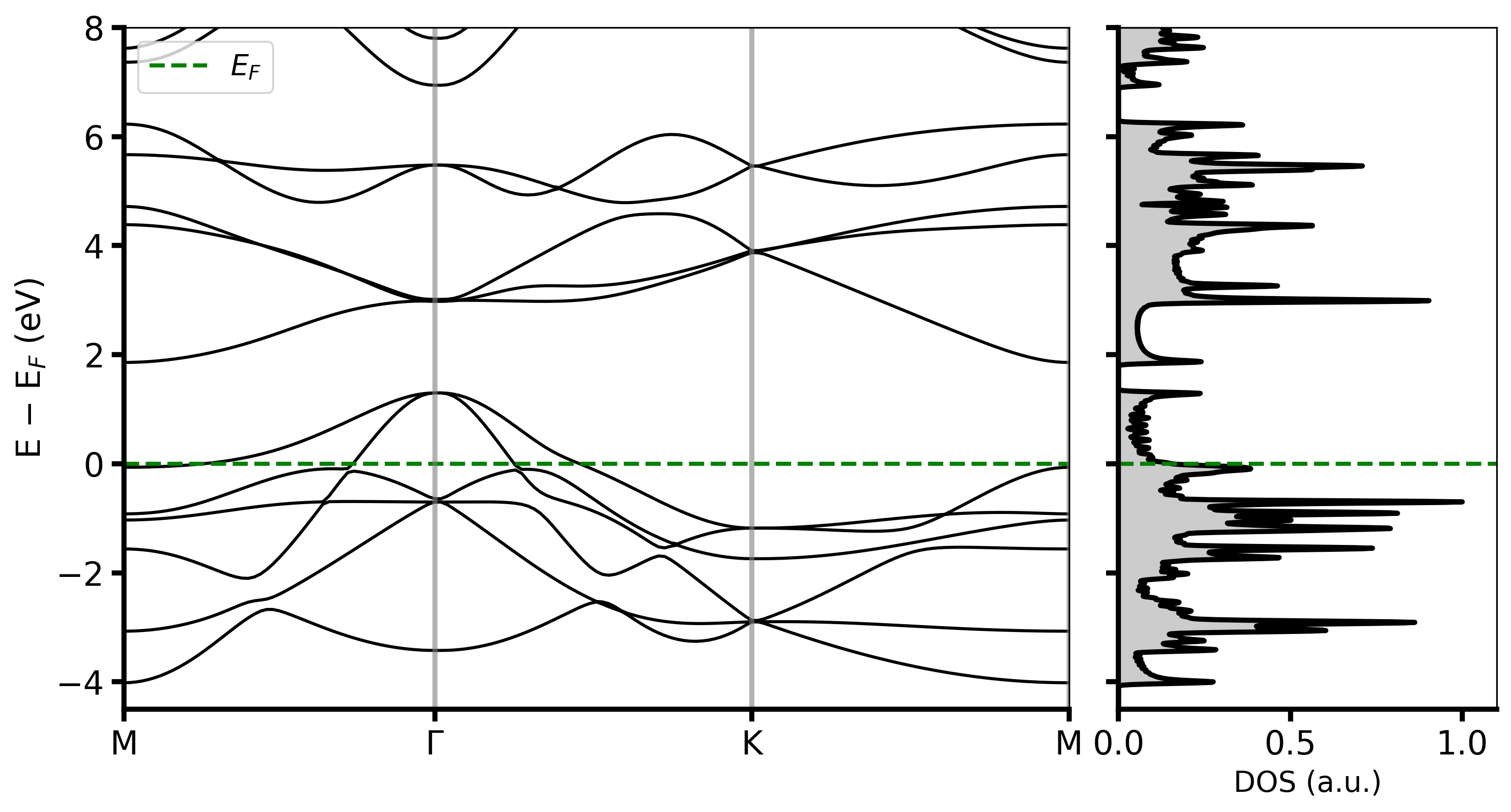}
\caption{PBE functional}
\end{subfigure}
\caption{Band structure and TDOS of the 1T phase of the {ScSe2} monolayer at a lattice parameter $a = \SI{3.52}{\angstrom}$, computed with (a) the LDA approximation and (b) the GGA-PBE functional.}
\label{fig:band1T}
\end{figure}

DFT calculations indicate a metallic character for the 1T phase with both the LDA and GGA-PBE functionals (Figure~\ref{fig:band1T}). Several bands cross the Fermi level ($E_F =$ \SI{-3.06}{eV} for LDA and \SI{-3.23}{eV} for PBE) along the $\Gamma$--$K$ and $M$--$\Gamma$ directions, confirming the absence of gap energy. This metallic character arises from the partial filling of the Sc $3d$ orbitals (electronic configuration $d^1$), which give rise to narrow bands near the Fermi level \cite{Chhowalla2013,Ma2011}.



\subsection{Band structure of the 2H phase}

The 2H phase exhibits trigonal-prismatic ($D_{3h}$) coordination at the metal site, with the primitive hexagonal unit cell containing one formula unit described by $\mathrm{Sc}(1/3,2/3,0.5)$, $\mathrm{Se}(2/3,1/3,+z)$, and $\mathrm{Se}(2/3,1/3,-z)$, with $z \approx 0.08$. The increased local symmetry partially lifts the degeneracy of the Sc $d$ orbitals, resulting in a more pronounced crystal-field splitting than in the 1T phase \cite{Cappelluti2013,Lv2015}.

The 2H band structure exhibits an indirect band gap between the $\Gamma$ and $K$ points of the Brillouin zone, with a computed value of approximately \SI{0.567}{eV} (LDA) and \SI{0.223}{eV} (GGA-PBE) (Table~\ref{tab:2H}).

\begin{table}[htbp]
\centering
\caption{Comparison of LDA and GGA-PBE results for the 2H phase at a lattice parameter $a = \SI{3.52}{\angstrom}$. }
\label{tab:2H}
\begin{tabular}{@{}lcccc@{}}
\toprule
Functional & $E_F$ (eV) & $E_g$ (eV) \\
\midrule
LDA & $-3.297$ & $0.567$ \\
GGA-PBE & $-2.785$ & $0.223$ \\
\bottomrule
\end{tabular}
\end{table}

\begin{figure}[h]
\centering
\begin{subfigure}[b]{0.75\textwidth}
\includegraphics[width=\textwidth]{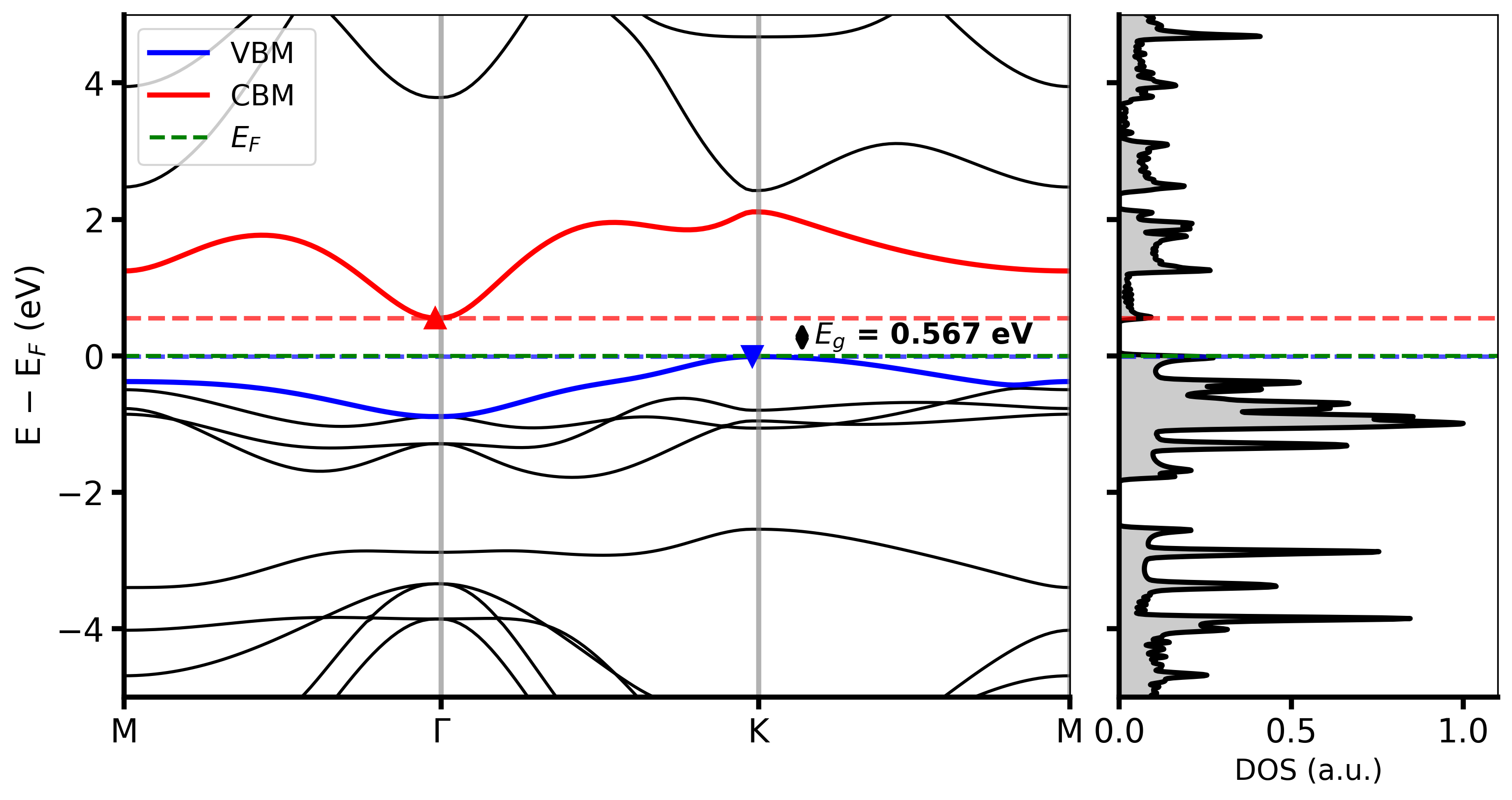}
\caption{LDA approximation}
\end{subfigure}\\ 
\begin{subfigure}[b]{0.75\textwidth}
\includegraphics[width=\textwidth]{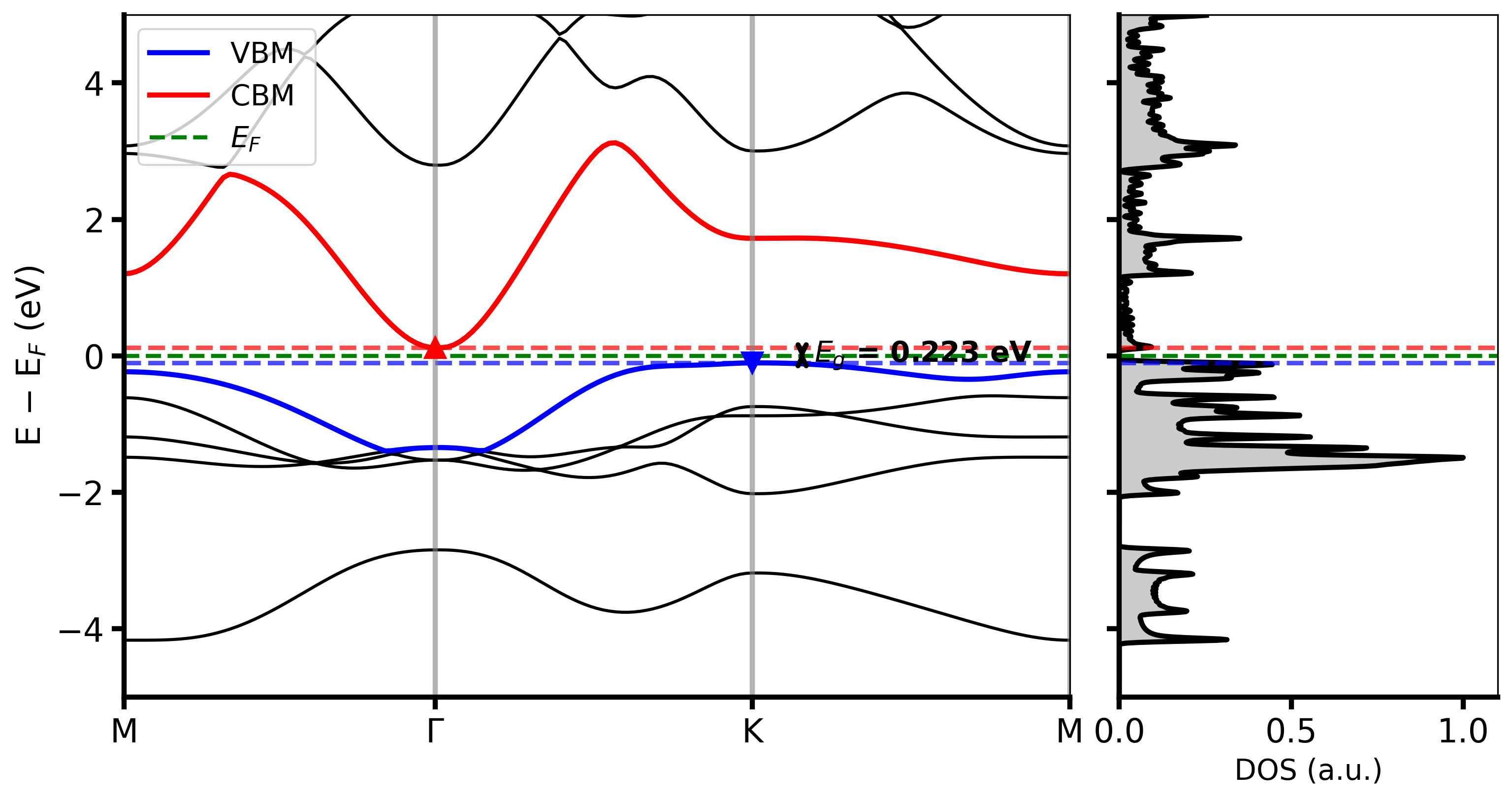}
\caption{PBE functional}
\end{subfigure}
\caption{Band structure and TDOS of the 2H phase of the {ScSe2} monolayer at a lattice parameter $a = \SI{3.52}{\angstrom}$, computed with (a) the LDA approximation and (b) the GGA-PBE functional.}
\label{fig:band2H}
\end{figure}

One can notice that with the \(\Gamma \)-point CB minimum  and \(K\)-point VB maximum, 2H phase establish an indirect band gap (Figure~\ref{fig:band2H}). This feature is particularly favorable for optoelectronic applications, since inter-band optical transitions in this case require phonon assistance \cite{Lv2015,Ramasubramaniam2012}. The upper VB shows significant dispersion along the $K$--$M$ direction, characteristic of strong Se$-4p$/Sc$-3d$ hybridization, while the lower CB, located essentially at the $K$ point, is dominated by Sc $d$ states \cite{Cappelluti2013,Dolui2013}.

The distinction between the two phases lies in their local coordination environment: the octahedral environment of the 1T phase favors the delocalization of $d$ electrons, enhancing the metallic character, whereas the prismatic environment of the 2H phase stabilizes the $d_{z^2}$ states below the Fermi level, opening a semiconducting gap. Total-energy calculations indicate that the 2H phase is more stable than the 1T phase by approximately \SI{0.8}{eV} per formula unit, consistent with previous studies of comparable systems \cite{Ataca2012,Tongay2014}.

\subsection{Comparison of results with LDA and GGA-PBE functionals}

The choice of exchange--correlation functional has a direct influence on the accuracy of the calculated electronic properties. Here we compare results obtained with the LDA, in the Perdew--Zunger parametrization \cite{PerdewZunger1981}, and with GGA-PBE, in the Perdew--Burke--Ernzerhof parametrization \cite{Perdew1996}, for both polymorphs of the {ScSe2} monolayer.

For the 1T phase, both functionals confirm the metallic character: several bands derived from the Sc$-3d$ orbitals cross the Fermi level along the $\Gamma$--$K$ and $K$--$M$ directions, consistent with a non-zero DOS at $E_F$ \cite{Kresse1996,Ma2011}. The lower VB, spanning approximately $-3$ to \SI{0}{eV}, is dominated by bonding {Se} $4p$ states, while the region between $0$ and \SI{0.5}{eV} is characterized by strong Sc$-3d$/Se$-4p$ hybridization, producing several nearly flat bands. This metallic behavior is consistent with the $d^1$ electronic configuration of Sc in an octahedral coordination environment \cite{Ataca2012,Wang2012}.

\begin{figure}[h]
\centering
\begin{subfigure}[b]{0.65\textwidth}
\includegraphics[width=\textwidth]{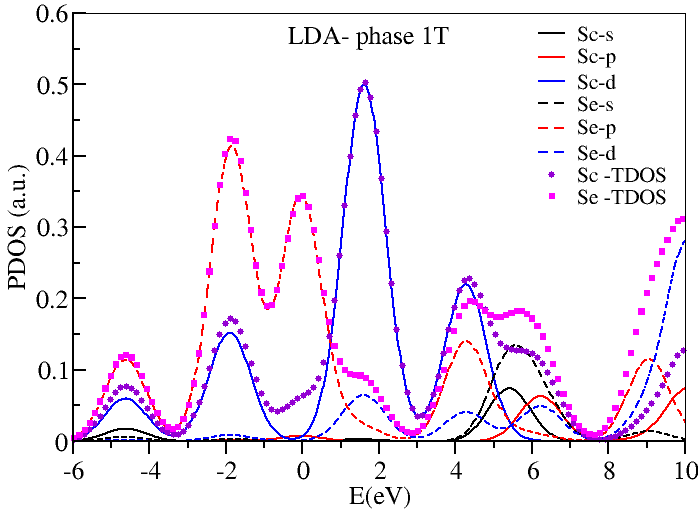}
\caption{LDA}
\end{subfigure}\\ 
\begin{subfigure}[b]{0.65\textwidth}
\includegraphics[width=\textwidth]{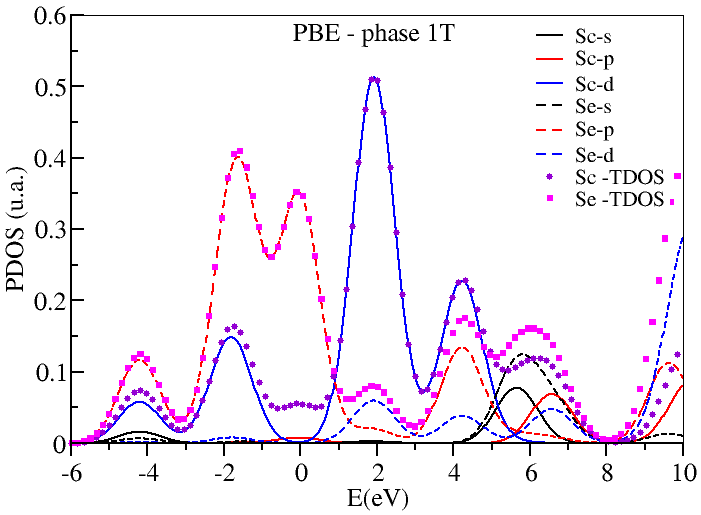}
\caption{PBE}
\end{subfigure}
\caption{PDOS as a function of energy for the 1T phase of the {ScSe2} monolayer at a lattice parameter $a = \SI{3.52}{\angstrom}$, obtained with the LDA approximation and GGA-PBE functional.}
\label{fig:pdos1T}
\end{figure}

For the 2H phase, the LDA functional yields an indirect gap of \SI{0.567}{eV}, with the CBM and VBM located at the $\Gamma $ and $K$ points, respectively, establishing a typical indirect gap consistent with the trigonal-prismatic $D_{3h}$ site symmetry \cite{Cappelluti2013,Mak2010}. As is well documented for LDA, this value is likely underestimated relative to more accurate methods, owing to the local character of the LDA exchange--correlation functional, which does not fully capture spatial density fluctuations in regions of low electron density \cite{PerdewZunger1981,Lebegue2009}. The VB spans approximately \SI{4.5}{eV}, with a pronounced DOS between $-4$ and \SI{0}{eV} originating from Sc$-3d$/{Se}$-4p$ hybrid states.
The systematic underestimation of the band gap by LDA relative to experimental or higher-level theoretical values is a well-documented limitation for 2D TMDs \cite{ChoTMDgap2012,Heyd2003}. 

The GGA-PBE functional partially corrects this deficit, increasing the 2H gap to \SI{0.223}{eV}, an increase of approximately 60\% relative to LDA, while preserving the indirect character of the gap between the $\Gamma$ and $K$ (equivalently labeled $M$ along the chosen path) symmetry points.

The qualitative picture obtained with GGA-PBE closely resembles that obtained with LDA: the deep {Se} $4p$ VBs lie between $-4$ and \SI{0}{eV}, while the Sc $d$ bands show wide dispersion near $E_F$. The lower CB lies slightly higher in energy than with LDA, consistent with the known tendency of PBE to yield somewhat larger lattice parameters and bandwidths \cite{Kresse1996,Eda2012}. The VB width computed with PBE is approximately 0.3--\SI{0.5}{eV} larger than with LDA \cite{Lebegue2009}, consistent with trends reported for {MoS2} \cite{ChoTMDgap2012} and {WSe2} \cite{Shi2013}, for which PBE systematically yields gaps 20--45\% larger than LDA.

\subsection{Density of States}

The DOS provides essential information on the distribution of electronic states as a function of energy, complementing the band-structure picture by integrating the contribution of all wave vectors $\mathbf{k}$ in the first Brillouin zone. We distinguish between the TDOS and the partial, or projected, PDOS, which resolves the contribution of each atomic species and angular-momentum channel ($s$, $p$, $d$, \ldots) \cite{Martin2004,Kresse1999}. The DOS is computed on a $12\times 12\times 1$ Monkhorst--Pack grid, using Gaussian smearing of \SI{0.05}{eV} for the metallic 1T phase and the tetrahedron method of Bl\"ochl for the semiconducting 2H phase \cite{Blochl1994}.


The TDOS of the 1T phase shows a non-zero value at the Fermi level, confirming the metallic character of this phase \cite{Chhowalla2013,Ma2011}. Three main spectral regions can be identified (see Figure \ref{fig:pdos1T}):
\begin{itemize}
\item 
{\bf Region I} ($-6.0$ to $-3.0$~eV): states predominantly of {Se}$-4p$ character, with partial Sc$-3d$ hybridization; this region corresponds to the lower VB and shows a relatively low TDOS, punctuated by a few Van Hove peaks.

\item 
{\bf Region II} ($-3.0$ to $0$~eV): strong Sc $3d$/{Se} $4p$ hybridization, characterized by a high TDOS reflecting the presence of several quasi-planar bands; most of the VB-states are located in this region.

\item 
{\bf Region III} ($0$ to $+3.0$~eV): CB-states governed by the unfilled Sc $d$ orbitals, with the TDOS increasing beyond $E_F$.
\end{itemize}

\begin{figure}[h]
\centering
\begin{subfigure}[b]{0.65\textwidth}
\includegraphics[width=\textwidth]{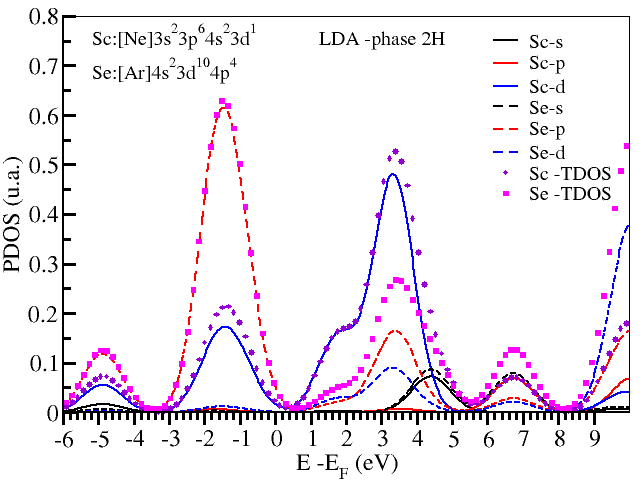}
\caption{LDA}
\end{subfigure} 
\begin{subfigure}[b]{0.65\textwidth}
\includegraphics[width=\textwidth]{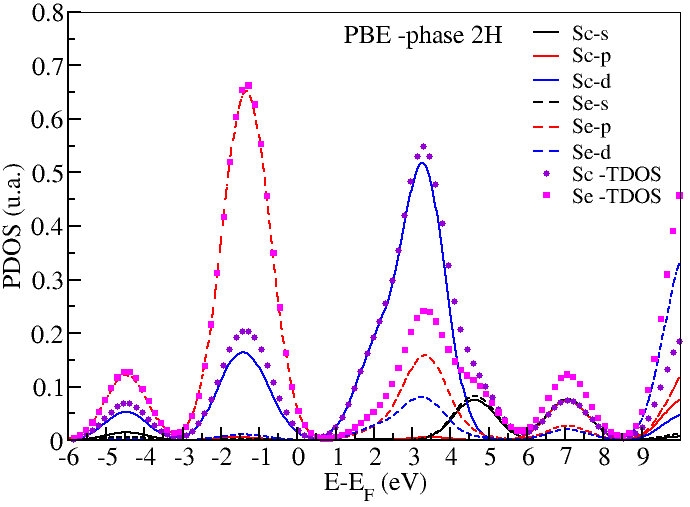}
\caption{PBE}
\end{subfigure}
\caption{PDOS as a function of energy for the 2H phase of the {ScSe2} monolayer at a lattice parameter $a = \SI{3.52}{\angstrom}$, obtained with the LDA and GGA-PBE functionals.}
\label{fig:pdos2H}
\end{figure}


The TDOS of the 2H phase shows a clear energy gap at the Fermi level, cleanly separating the VB and CB \cite{Lv2015,Dolui2013}. Three main spectral regions can again be identified (see Figure \ref{fig:pdos2H}):
\begin{itemize}
\item 
{\bf Region I} ($-6.0$ to $-3.0$~eV): states dominated by {Se}$-4p$ orbitals, with a minor Sc$-3d$ contribution; this region forms the lower part of the VB and is structured by a few Van Hove singularities.

\item 
{\bf Region II} ($-3.0$ to $0$~eV): dominated by a pronounced peak centered near $-1.5$~eV, primarily of {Se}$-4p$ origin, with significant Sc$-3d$/{Se}$-4p$ hybridization; this peak constitutes the top of the VB and reflects strongly flattened bands, where the TDOS reaches its maximum value in the VB region.

\item 
{\bf Region III} ($0$ to $+5.0$~eV): the CB, governed by the unfilled Sc $3d$ orbitals, exhibits two main features: a first peak near $+3.0$~eV of predominantly Sc $3d$ character, and a second, broader peak near $+3.5$ to $+4.0$~eV. A clear gap separates the VB and CB, confirming the semiconducting character of the 2H phase.
\end{itemize}

Unlike the 1T phase, the DOS at the Fermi level, $N(E_F)$, vanishes for the 2H phase, consistent with the existence of a semiconducting gap of $E_g \approx \SI{0.567}{eV}$ (LDA) and $E_g \approx \SI{0.223}{eV}$ (PBE), with the CBM at $\Gamma$ and the VBM at $K$ \cite{Chhowalla2013}.
The PDOS is obtained by projecting the Kohn--Sham wavefunctions onto atom-centered orbitals \cite{Kresse1999,Bader1990}, allowing a quantitative assessment of each orbital's contribution to the various spectral regions. The Sc atom ($Z=21$, configuration [Ne]$3s^2 3p^6 3d^1 4s^2$) contributes primarily through its $3d$ orbitals; the lifting of their degeneracy by the crystal field (octahedral in the 1T phase, prismatic in the 2H phase) produces distinct PDOS profiles for the two polymorphs, as illustrated in Figures~\ref{fig:pdos1T} and~\ref{fig:pdos2H}. The extended range of the Sc $d$-orbital PDOS (approximately 4--\SI{6}{eV}) suggests significant hybridization with the {Se} $p$ orbitals, consistent with bond-chemistry arguments and Bader charge analysis \cite{Bader1990}.

\section{Dielectric and Optical Properties}
\label{sec:optical}

The complex dielectric function,
\begin{equation}
\varepsilon(\omega) = \varepsilon_1(\omega) + i\,\varepsilon_2(\omega),
\label{eq:eps}
\end{equation}
is one of the essential quantities governing the optical and electronic response of a material to an external electromagnetic field. For two-dimensional TMDs, $\varepsilon(\omega)$ provides a direct connection to optical absorption processes, interband electronic transitions, and intrinsic plasmonic properties \cite{Dressel2002,Cudazzo2011}. The {ScSe2} monolayer is particularly interesting in this respect, given the coexistence of two crystalline polymorphs --- the 2H phase (prismatic coordination, $D_{3h}$ point group) and the 1T phase (octahedral coordination, $D_{3d}$ point group) --- which give rise to intrinsically distinct electronic, and consequently dielectric, structures, offering phase polymorphism as a control parameter for the optical response \cite{Ataca2012,Guo2014}.

The dielectric function was computed within linear-response theory, 
using the PBE-GGA functional with van der Waals corrections, as implemented in CP2K \cite{Dressel2002,Guo2014}. The imaginary part of eq.(\ref{eq:eps}) is obtained as a sum over all allowed inter-band transitions between VB and CB states,
\begin{equation}
\varepsilon_2(\omega) \;\propto\; \sum_{c,v,\mathbf{k}} \left| \langle \psi_c | \hat{p} | \psi_v \rangle \right|^2 \delta\!\left(E_c - E_v - \hbar\omega\right),
\label{eq:eps2}
\end{equation}
where $\psi_c$ and $\psi_v$ denote the CB and VB wavefunctions, $\hat{p}$ is the momentum operator, and the Dirac delta function enforces energy conservation. The peaks of $\varepsilon_2(\omega)$ directly identify the energies of the active optical transitions \cite{Dressel2002,Guo2014}. The real part of eq.(\ref{eq:eps}) is obtained from $\varepsilon_2(\omega)$ through the Kramers--Kronig relation,
\begin{equation}
\varepsilon_1(\omega) = 1 + \frac{2}{\pi}\, \mathcal{P}\!\!\int_0^{\infty} \frac{\omega' \varepsilon_2(\omega')}{\omega'^2 - \omega^2}\, d\omega',
\label{eq:kk}
\end{equation}
which guarantees causality of the system's response.

For a non-magnetic material such as {ScSe2} ($\mu_r \approx 1$), the complex refractive index $\tilde{n}(\omega) = n(\omega) + i\kappa(\omega)$ is related to the dielectric function through $\tilde{n}^2(\omega) = \varepsilon(\omega)$, which gives
\begin{align}
n(\omega) &= \sqrt{\frac{|\varepsilon(\omega)| + \varepsilon_1(\omega)}{2}}, &
\kappa(\omega) &= \sqrt{\frac{|\varepsilon(\omega)| - \varepsilon_1(\omega)}{2}},
\label{eq:nk}
\end{align}
with $|\varepsilon(\omega)| = \sqrt{\varepsilon_1^2(\omega) + \varepsilon_2^2(\omega)}$. The absorption coefficient and the normal-incidence reflectivity follow from the Fresnel relations,
\begin{equation}
\alpha(\omega) = \frac{2\omega}{c}\,\kappa(\omega), \qquad
R(\omega) = \frac{\left(n(\omega)-1\right)^2 + \kappa^2(\omega)}{\left(n(\omega)+1\right)^2 + \kappa^2(\omega)}.
\label{eq:alphaR}
\end{equation}

\begin{figure*}[htbp]
\centering
\includegraphics[width=1.0\textwidth]{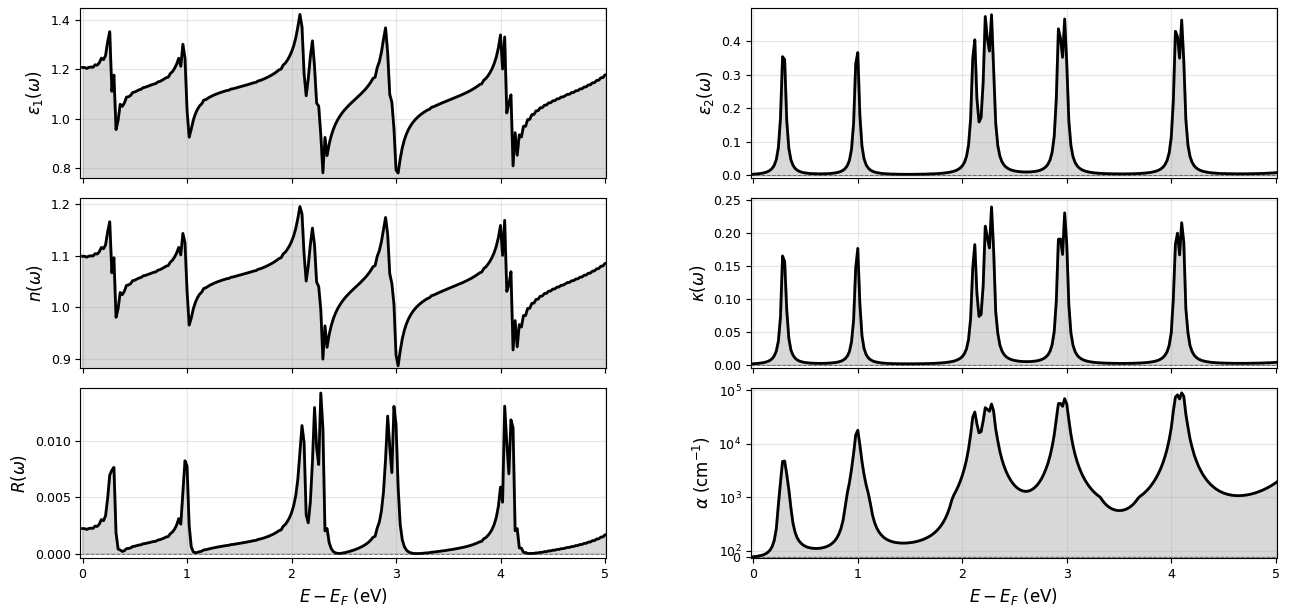}
\caption{Real part $\varepsilon_1(\omega)$, imaginary part $\varepsilon_2(\omega)$, refractive index $n(\omega)$, extinction coefficient $\kappa(\omega)$, reflectivity $R(\omega)$, and absorption coefficient $\alpha(\omega)$ of the 2H phase of the {ScSe2} monolayer at $a = \SI{3.52}{\angstrom}$.}
\label{fig:opt2H}
\end{figure*}

\begin{figure*}[htbp]
\centering
\includegraphics[width=1.0\textwidth]{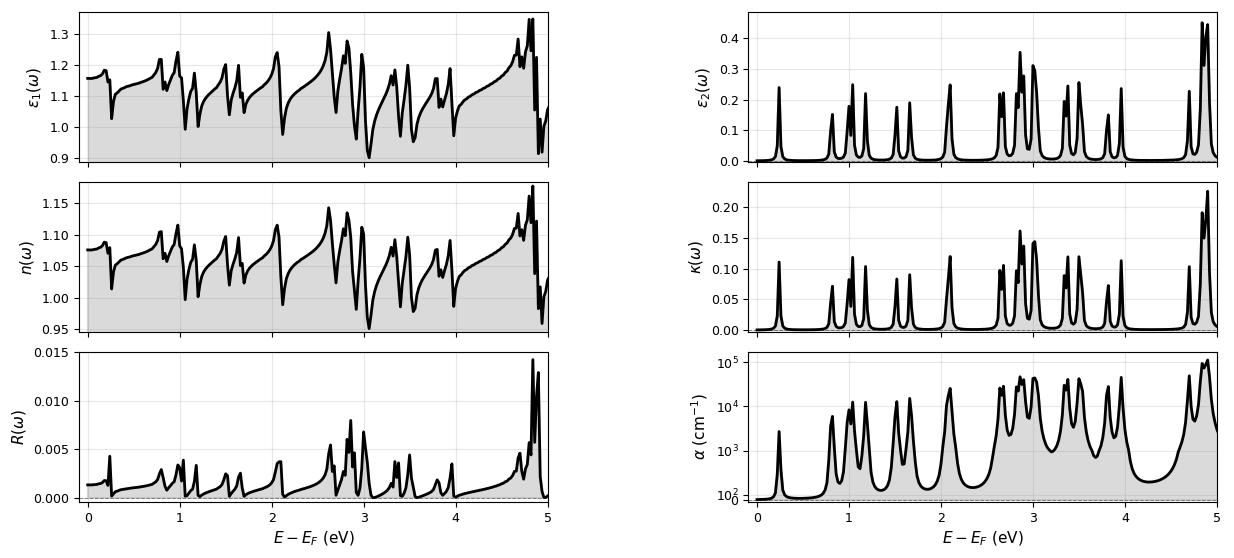}
\caption{Real part $\varepsilon_1(\omega)$, imaginary part $\varepsilon_2(\omega)$, refractive index $n(\omega)$, extinction coefficient $\kappa(\omega)$, reflectivity $R(\omega)$, and absorption coefficient $\alpha(\omega)$ of the 1T phase of the {ScSe2} monolayer at $a = \SI{3.52}{\angstrom}$.}
\label{fig:opt1T}
\end{figure*}

The optical spectra of the {ScSe2} monolayer, computed for both polymorphs at the same lattice parameter $a = \SI{3.52}{\angstrom}$, are presented in Figures~\ref{fig:opt2H} and~\ref{fig:opt1T}. Each figure groups the five optical quantities obtained via the joint-density-of-states (JDOS) method combined with the exact Kramers--Kronig transform: the real and imaginary parts of the dielectric function, $\varepsilon_1(\omega)$ and $\varepsilon_2(\omega)$; the refractive index $n(\omega)$; the extinction coefficient $\kappa(\omega)$; the normal-incidence reflectivity $R(\omega)$; and the absorption coefficient $\alpha(\omega)$, over the spectral window 0--\SI{5}{eV}. The \SI{20}{\angstrom} vacuum spacing in the supercell ensures the electrostatic isolation of the monolayer from its periodic replicas \cite{Dronskowski1993}.

For both phases, our GGA-PBE calculations yield a static dielectric constant $\varepsilon_1(0) \approx 1.30$(2H)--$1.16$(1T), reflecting the low-frequency regime in which electronic polarization is not yet activated by interband transitions. 
The phases begin to diverge only near the onset of optical absorption: the first resonance of the 1T phase occurs at approximately \SI{0.24}{eV}, compared to \SI{0.30}{eV} for the 2H phase. The proximity of these thresholds indicates that the optical gaps of the two {ScSe2} polymorphs are comparable, although the 1T phase exhibits a slightly larger $\varepsilon_1$ resonance amplitude (maximum of 1.2, versus 1.6 for the 2H phase), consistent with a somewhat stronger electronic polarizability in the octahedral coordination.

The 1T phase displays a markedly different spectral response: rather than five regularly spaced peaks, $\varepsilon_2(\omega)$ organizes into compact clusters of several transitions (see Figure~\ref{fig:opt1T}). 
This distribution reflects the splitting of the Sc $d$ orbitals into $t_{2g}$/$e_g$ levels under octahedral ($D_{3d}$) coordination, which groups several interband transitions within narrow energy ranges rather than distributing them across the full spectral window.

The two-dimensional nature of the {ScSe2} monolayer induces an intrinsic dielectric anisotropy: the in-plane ($\varepsilon_{xx} = \varepsilon_{yy}$) and out-of-plane ($\varepsilon_{zz}$) components of $\varepsilon(\omega)$ diverge significantly, even at low frequency \cite{Cudazzo2011,Zhao2021}. Along the out-of-plane direction, quantum confinement leads to a renormalization of excitonic effects: in the absence of a surrounding dielectric medium, the reduced electrostatic screening increases the exciton binding energy to several hundred meV, as observed in other group-IV TMDs \cite{Zhao2021}. This effect constitutes a fundamental difference between the optical response of the monolayer and that of the corresponding bulk crystal.

\section{Conclusion}

We have investigated the {ScSe2} monolayer in its two polymorphic forms, 1T (octahedral, $D_{3d}$) and 2H (trigonal prismatic, $D_{3h}$), through their band structure, DOS, 
and optical/dielectric properties. The results show that the local coordination of scandium, through the lifting of degeneracy of its $3d$ orbitals, governs all the properties examined.

The 1T phase is metallic, with Sc$-3d$ bands crossing $E_F$ in both LDA and PBE, while the 2H phase exhibits a robust indirect gap ($\SI{0.567}{eV}$ with LDA, $\SI{0.223}{eV}$ with GGA-PBE) between $\Gamma$ and $K$. PBE systematically shifts the LDA gap upward (by approximately 60\% for the 2H phase) and, in some cases, can qualitatively alter the predicted electronic character of the system.
Both phases have VBs dominated by {Se}$-4p$ states but differ near the Fermi level: in the 1T phase, the partially filled $t_{2g}/e_g$ levels give rise to a non-zero DOS at $E_F$ ($N(E_F) \neq 0$), whereas in the 2H phase the $d_{z^2}$ orbital, stabilized below $E_F$, opens a well-defined gap ($N(E_F) = 0$). 

For both phases, $\varepsilon_1(0)$ and the absorption onset (approximately 0.24--\SI{0.30}{eV}) are similar. However, their spectral profiles differ markedly: Five $\varepsilon_2(\omega)$ peaks distributed between 0.3 and \SI{5.0}{eV} for the 2H phase ($D_{3h}$ symmetry), versus densely grouped separated by a transparency window for the 1T phase. Compared to {MoS2}, {WS2}, {TiSe2}, {ZrSe2}, and {HfSe2}, {ScSe2} combines a semiconducting polytype with an intermediate band gap (2H) and a structured metallic polytype (1T) within a single chemical composition.

The finding emerge from this work:
a metal-to-semiconductor transition controlled solely by the 1T/2H polymorphism, with no change in chemical composition;
beneficial for band-gap engineering;
and markedly distinct optical signatures between the two phases, potentially useful for energy-selective optoelectronic devices.

\bmhead{Acknowledgements}

The authors gratefully acknowledge the Advanced Materials and Quantum Phenomena Laboratory, Faculty of Sciences of Tunis, Tunis El Manar University, for providing the computational resources used in this work.

\section*{Declarations}


\begin{itemize}
\item No Funding
\item The authors declare that they have no known competing financial interests or personal relationships that could have appeared to influence the work reported in this paper.
\item No restrictions on data created or analyzed as part of this study.
\item Authors contribution: All authors contributed to the study conception and design. Material preparation, data collection and analysis were performed by all authors. The first draft of the manuscript was written by Mohsen Jemai, and all authors commented on previous versions of the manuscript. All authors read and approved the final manuscript.
\end{itemize}


\end{document}